\documentclass{jfm}
\usepackage{graphicx}
\usepackage[table]{xcolor}
\usepackage{colortbl,tikz,amsmath}

\usepackage{epstopdf, epsfig}
\usepackage[sc]{caption}
\usepackage{subcaption}
\usepackage[colorlinks = true,
            linkcolor = black,
            urlcolor  = black,
            citecolor = black,
            anchorcolor = black]{hyperref}

\shorttitle{Effects of fetch length on boundary layer recovery}
\shortauthor{M. Formichetti, D. D. Wangsawijaya, S. Symon, B. Ganapathisubramani}

\title{Effects of fetch length on turbulent boundary layer recovery past a \\ step-change in surface roughness}

\author{Martina Formichetti\aff{1}
  \corresp{\email{martina.formichetti@soton.ac.uk}},
  Dea D. Wangsawijaya\aff{1}, \\
  Sean Symon\aff{1},
 \and Bharathram Ganapathisubramani\aff{1}}

\affiliation{\aff{1}Department of Aeronautical and Astronautical Engineering, University of Southampton, University Road, Southampton, SO17 1BJ, UK}

\begin{document}

\maketitle

\begin{abstract}
        Recent studies focusing on the response of turbulent boundary layers (TBL) to a step-change in roughness have provided insight into the scaling and characterisation of TBLs and the development of the internal layer. Although various step-change combinations have been investigated, ranging from smooth-to-rough to rough-to-smooth, the ``minimum" required roughness fetch length over which the TBL returns to its homogeneously rough behaviour remains unclear. Moreover, the relationship between a finite- and infinite-fetch roughness function (and the equivalent sandgrain roughness) is also unknown. In this study, we determine the minimum ``equilibrium fetch length" for TBL developing over a smooth-to-rough step-change as well as the expected error in local skin friction if the fetch length is under this minimum threshold. An experimental study is carried out where the flow is initially developed over a smooth wall, and then a step-change is introduced using patches of P24 sandpaper. 12 roughness fetch lengths are tested in this study, systematically increasing from $L = 1\delta_2$ up to $L = 39\delta_2$ (where \textit{L} is the roughness fetch length and $\delta_2$ is the TBL thickness of the longest fetch case), measured over a range of Reynolds numbers ($4\cdot10^2 \leq Re_\tau \leq 2\cdot10^5$). Results show that the minimum fetch length needed to achieve full equilibrium recovery is around $20\delta_2$. Furthermore, we observe that $C_f$ recovers to within 10\% of its recovered value for fetch lengths $\geq 5\delta_2$. This information allows us to incorporate the effects of roughness fetch length on the skin friction and roughness function.
\end{abstract}


\section{Introduction}
\noindent Turbulent Boundary Layers (TBLs)  developing over rough walls encompass many engineering applications. Studying this phenomenon is crucial for the performance evaluation of an engineering system. For example, in the aeronautical or automotive industry, the manipulation of a TBL using a surface treatment (i.e. ``roughness") may result in drag reduction, \citep{Whitmore2002}. On the other hand, in the wind energy sector, an Atmospheric Boundary Layer (ABL) in neutral conditions developing over a wind farm behaves like a large-scale TBL over ``roughness". Understanding the physics of this flow leads to more accurate wind power predictions and strategic site selections, \citep{BouZeid2020}. 

A realistic representation of a rough-wall TBL in these applications hardly ever involves a ``homogeneous" rough wall. In some scenarios, it can be better approximated with a streamwise transition in roughness. For example, the roughness on a ship hull (due to biofouling and coating deterioration) incurs at various roughness length scales and sites, resulting in multiple streamwise transitions in roughness, affecting the TBL developing over it. At the same time, when analysing sites for wind farm locations we might encounter areas of complex terrain where we see a combination of forests and plains or sea and coastline. These variations significantly affect the behaviour of the ABL and, consequently, the drag production near the surface.

\begin{figure}
\centerline{\includegraphics{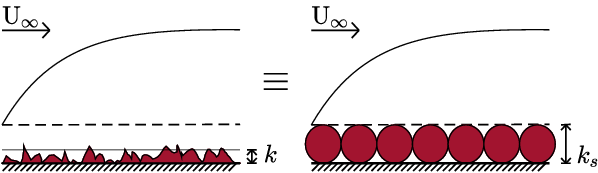}}
\captionsetup{width=.95\linewidth}
\caption{Schematic of physical roughness height $k$ vs. the equivalent sand-grain roughness height $k_s$.}
\label{fig1} 
\end{figure}

The main change occurring in a TBL over a rough wall compared to one over a smooth wall is an increase in Wall Shear Stress (WSS) and, consequently, a momentum deficit $\Delta U^+$, characterised by a vertical shift in the logarithmic layer of the streamwise mean velocity profile, which, for fully rough flows, is defined as follows:
\begin{equation}
U^+=\frac{1}{\kappa}ln\left(y^+\right)+B-\Delta U^+ = \frac{1}{\kappa}ln\left( \frac{y}{k_s}\right)+B_{FR},
\label{eq:velocitylaw}
\end{equation}
where $\kappa\approx 0.39, B\approx4.3, \text{ and } B_{FR}=8.5$. Equation \ref{eq:velocitylaw} shows that the main two parameters used to scale TBLs over rough walls are $k_s$ as the length scale, and the friction velocity $u_\tau$ (see \citet{Jimenez2004} or other similar works for the details on the scaling arguments). A surface with arbitrary representative roughness height $k$ is associated with a length scale $k_s$, as shown in figure \ref{fig1}, which affects the logarithmic layer of the mean velocity profile in the same way as a surface covered by an ideally uniform sand-grain type of roughness with physical height $k_s$. Its definition is given in \citet{Colebrook1937} and \citet{Nikuradse1933} and some examples of its usage can be found in \citet{Flack2014} and \citet{Schultz2009}. This height is usually calculated by taking a point measurement in the logarithmic layer of a TBL and using equation \ref{eq:velocitylaw}, with the main assumption being that the flow is within the fully rough regime. Another method of calculating $k_s$ is given by \citet{Monty2016}, which consists of an iterative procedure to obtain a direct relation between the surface friction and $k_s$. This method assumes that the TBL adheres with the outer layer similarity (see \citet{Townsend1965}) thus, the TBL is in equilibrium with the surface texture.

The response of the WSS after an abrupt change in roughness and its recovery to an equilibrium state has been studied extensively, using both experimental techniques and numerical simulations. The main results found in research are that the WSS either increases or decreases abruptly overshooting or undershooting the expected value for the downstream surface in smooth-to-rough \citep{Antonia1971,Bradley1968} and rough-to-smooth transitions \citep{Antonia1972,Efros2011,Hanson2016}, respectively. Experimentally, this has been researched with direct WSS measurements immediately downstream of the step change by using floating element balances, \citep{Bradley1968,Efros2011}, near-wall hot-wires, \citep{Chamorro2009}, Preston tubes, \citep{Louriero2010}, and pressure taps, \citep{Antonia1971,Antonia1972}, coupled with indirect methods to obtain the development of the WSS with distance from the step-change. This was mainly done using a logarithmic fit to match the measured value and the expected one for the downstream surface if there were no surface changes upstream. Numerically, the WSS recovery after a step-change in roughness has been mainly investigated with DNS \citep{Lee2015,Ismail2018,Rouhi2019} and LES \citep{Saito2014,Sridhar2018}. The results of all these studies were conducted at different Reynolds numbers, approximately $10^2\leq Re_\tau\leq 10^6$, and a variety of upstream-to-downstream roughness height ratios, $-6 \leq ln(z_{02}/z_{01})\leq 6$ (where negative ratios correspond to rough-to-smooth transitions, and positive values correspond to smooth-to-rough changes). 

Previous studies highlighted some remaining questions regarding the TBL recovery to an equilibrium condition after being subjected to a streamwise step change in roughness. Firstly, as mentioned above, the characteristic overshoot or undershoot of the WSS just after a step-change in roughness renders the characterisation methods developed for the homogeneous rough wall inaccurate, since both scaling parameters depend on WSS and are calculated assuming fully rough homogeneous roughness. This leads to a need to define a minimum recovery length in which the flow recovers to the homogeneous rough wall TBL. Secondly, the use of experimental indirect methods and numerical methods to obtain the WSS recovery after a step change resulted in a wide range of recovery fetch lengths between $1\delta$ and $10\delta$, making it difficult to draw specific conclusions from these predictions. Moreover, some studies such as \citet{Saito2014, Sridhar2018} showed an increase in recovery fetch length with Reynolds number which is inconsistent with other studies, highlighting the necessity of a direct WSS measurement method for a more accurate prediction. An extensive review and comparison between existing studies can be found in \citet{Li2020}.

In this study, we consider a TBL developing from a baseline smooth wall and subjected to a streamwise transition to a rough wall. We aim to investigate and achieve a reliable value for the minimum roughness fetch length that allows a TBL developing past such step change in roughness to recover to an equilibrium condition, i.e. fully adjust to the rough wall downstream of the transition. This is essential since all of the scaling arguments used in rough-wall TBLs depend on the WSS and the latter changes drastically after a step-change in wall roughness. Secondly, we aim to quantify the error in choosing a shorter fetch to conduct experiments/simulations on presumably homogeneous fully rough flows. This would be helpful to quantify the uncertainty of the data if, for instance, a study needed to be conducted in a facility with a shorter test section, or if there were limitations on the domain size for a numerical investigation dictated by the available computational power. Finally, we aim to develop a relationship between the $k_s$ of a surface with short fetch (where the flow is not in equilibrium) in terms of the equilibrium value of $k_s$. To answer these questions, we designed an experiment to directly measure the change in WSS to sequential increases in roughness fetch, covering a wide range of Reynolds number, $4\cdot 10^3\leq Re_\tau \leq 2\cdot 10^5$, to ensure all or most common conditions are covered. The setup of the experiment is covered in \S \ref{methods}, followed by a detailed discussion of the results in \S \ref{results} leading to the conclusions and future work in \S \ref{conclusion}.

\section{Experimental set-up and methodology}
\label{methods}
\begin{figure}
\centering
\begin{subfigure}{0.5\textwidth}
\centerline{\includegraphics{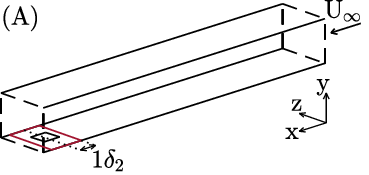}}
\captionsetup{width=.95\linewidth}
\end{subfigure}\begin{subfigure}{0.5\textwidth}
\centerline{\includegraphics{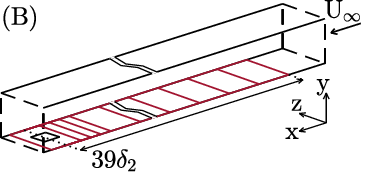}}
\captionsetup{width=.95\linewidth}
\end{subfigure}
\caption{Schematic of the experiment with the fetch length, $L$, measured from the centreline of the balance: (A) $L=1\delta_2$ and (B) $L=39\delta_2$ }
\label{fig2}
\end{figure}

\noindent The experimental campaign is conducted inside the closed return BLWT at the University of Southampton. The TBL is tripped by a turbulator tape located at the inlet of the test section, marking the streamwise datum (x=0) and further developed along the floor of the 12 m-long test section, which has a width and height of 1.2$ \times $1 m. Figure \ref{fig2} illustrates the tunnel and coordinate system where x, y and z denote the streamwise, wall-normal and spanwise directions, respectively. The tunnel is equipped with a “cooling unit” comprised of two heat exchangers and a temperature controller such that the air temperature remains constant during measurements ($21^{\circ}C \pm 0.5^{\circ}C$). The free stream has a turbulence intensity of $\leq 0.1 U_\infty$ (where $U_\infty$ is the free-stream velocity), measured with hot-wire anemometry before the experimental campaign. The tunnel is equipped with a closed-loop PID controller to set $U_\infty$, and air properties are measured with a pitot-static tube and a thermistor inside the BLWT.

As seen in figure \ref{fig2}, the experiment consisted of a roll of P24 sandpaper cut in patches of size $2\delta_2\times 8\delta_2$ (where $\delta_2$ refers to the TBL thickness of the case with the longest fetch measured at the balance location). The patches are sequentially taped on the floor of the wind tunnel's test section starting at the measurement point, about 59delta2 downstream from the test section's inlet, and added upstream. In this way, the roughness fetch measured from the centre-line of the balance is systematically increased, with the shortest fetch being $1\delta_2$, and the longest being $39\delta_2$. All cases are listed in table \ref{tb:legend}. The longest fetch tested was chosen as a threshold between having as long of a fetch as achievable in our facility and ensuring the TBL on the smooth surface upstream of the step change would also have enough development length to be in equilibrium conditions ($\approx25\delta_1$ in the longest fetch case, where $\delta_1$ is the TBL thickness of the smooth surface measured at the measurement point).

The experimental campaign was designed to take direct WSS measurements at different Reynolds numbers and with sequentially increased roughness fetch (the distance between the step change in roughness and the measurement point). This was possible by employing a floating element drag balance (located at the previously mentioned ``measurement point"), designed and manufactured by \citet{Ferreira2024}. With this tool, the friction on the wall was monitored during velocity sweeps (0-40 [ms$^{-1}$]) while systematically increasing the length of the roughness fetch. The velocity sweeps were run three times per case to ensure the repeatability of the results. A schematic of the balance and its specifications can be found in \citet{Ferreira2024}. 

For each fetch length, measurements are conducted within a range of freestream velocities $10\leq U_\infty \leq40\;\; \text{[ms$^{-1}$]}$, allowing 10 seconds for the flow to adjust after each velocity increase and 60 seconds for the flow to come to rest completely before restarting the sweep. The sampling rate was set to $f_s = 256 Hz$, while the sampling time was set to 60s. Once the friction force, $F$, has been measured, the WSS, $\tau_w$, and friction velocity, $u_\tau$, can be directly computed.
\begin{equation}
    \tau_w = \frac{F}{A} = \frac{1}{2}\rho U_\infty^2 C_f = \rho u_\tau^2 ,
    \label{eq:utau}
\end{equation}
where $A$ is the surface area of the balance plate and $C_f$ is the friction coefficient.

Planar PIV was also performed in the streamwise wall-normal plane above the floating element location. This was done to check whether the \textit{Outer Layer Similarity} used in \citet{Monty2016} to calculate $k_s$ holds for some or all of our cases and to calculate the boundary layer thickness for all cases. The additional PIV measurements were only conducted at a free-stream velocity of $20 \text{[ms$^{-1}$]}$ and only for the cases with fetch length $1\delta_2,3\delta_2,5\delta_2,7\delta_2,9\delta_2,19\delta_2, \text{ and }39\delta_2$. The selection of free-stream velocity and fetches to study with PIV was dictated by the trends obtained in the drag balance measurements, as seen in \S \ref{results}. The data was sampled at $f_s =1 Hz$ ($t_r=U_\infty/(f_s\cdot \delta_2)\approx133$, where $t_r$ is the TBL turnover rate) with Lavision Imager CMOS 25 MP cameras (resolution of 17 pixels/mm), using a Bernoulli 200 mJ, 532 nm Nd:YAG laser and the Lavision software Davis 10 for acquisition. The data was processed using an in-house code for cross-correlation, with a final window size of 16$\times$16 px with 75\% overlap, and a viscous-scaled final window size of 30$\times$30 [cm]. 

\section{Results}
\label{results}
\begin{figure}
\begin{subfigure}{0.5\textwidth}
\centerline{\includegraphics{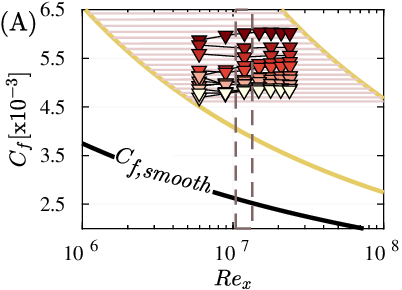}}
\end{subfigure}\begin{subfigure}{0.5\textwidth}
\centerline{\includegraphics{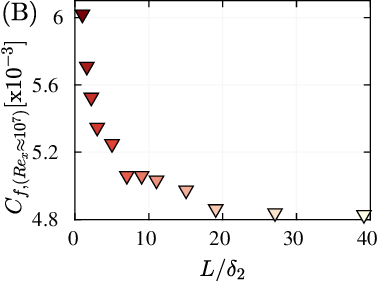}}
\end{subfigure} 
\caption{In (A): $C_f$ plotted against $Re_x=U_\infty x/\nu$, where $x$ is the distance of the balance centreline from the test-section's inlet, with (\textcolor[rgb]{0.9,0.8,0.4} {\huge{\textbf{-}}}) being lines of unit $Re=U_\infty/\nu$, and (\textcolor[rgb]{0.9,0.8,0.8} {\huge{\textbf{-}}}) lines of constant $k_s/x$ as described by \cite{Monty2016}. In (B)  $C_f$ at $Re_x\approx10^7$ plotted against the fetch length $L$ normalised by the downstream TBL thickness $\delta_2$}
\label{fig3}
 \end{figure}

\noindent The evolution of the friction coefficient obtained with the drag balance at different fetch lengths and increasing Reynolds number ($Re_x=\rho U_\infty x/\mu$, where $x$ is the streamwise distance between the wind-tunnel's test section inlet and the balance centre-line) can be seen in figure \ref{fig3}A. The colour legend for all the plots in \S\ref{results} is shown in table \ref{tb:legend}, while the error propagation given by the drag balance measurements is listed in table \ref{tab:unc} for the different parameters. Figure \ref{fig3}A shows that for a fixed fetch length, $C_f$ is independent of $Re_x$, which is a sign that the flow is within the fully rough regime bounds mentioned in \S1. However, it is not fetch-length independent: the fetch length is inversely proportional to $C_f$, consistent with the overshoot downstream of the transition observed by multiple studies listed in \citet{Li2020}, and the slow recovery with increasing distance from the step change.
\definecolor{col1}{rgb}{0.468750000000000	0	0.0312500000000000}
\definecolor{col2}{rgb}{0.647879464285714	0.110491071428571	0.104910714285714}
\definecolor{col3}{rgb}{0.767299107142857	0.184151785714286	0.154017857142857}
\definecolor{col4}{rgb}{0.827008928571429	0.220982142857143	0.178571428571429}
\definecolor{col5}{rgb}{0.886718750000000	0.257812500000000	0.203125000000000}
\definecolor{col6}{rgb}{0.902901785714286	0.363839285714286	0.302455357142857}
\definecolor{col7}{rgb}{0.919084821428571	0.469866071428572	0.401785714285714}
\definecolor{col8}{rgb}{0.935267857142857	0.575892857142857	0.501116071428571}
\definecolor{col9}{rgb}{0.951450892857143	0.681919642857143	0.600446428571429}
\definecolor{col10}{rgb}{0.967633928571429	0.787946428571429	0.699776785714286}
\definecolor{col11}{rgb}{0.983816964285714	0.893973214285714	0.799107142857143}
\definecolor{col12}{rgb}{1,1,0.898437500000000}

\begin{table}
    \centering
    \begin{tabular}{|c|c|c|c|c|c|}
    \hline
    $C_f$ & $u_\tau\;\text{[ms$^{-1}$]}$  & $Re_x$ & $\delta \; [m]$ & $U \;\text{[ms$^{-1}$]}$ & $k_s \; [m]$\\
    \hline
    $\pm 8.6\cdot10^{-6}$ & $\pm 3.3 \cdot10^{-3}$ &
         $\pm 3.2\cdot 10^3$ & $\pm 1\cdot10^{-2}$ & $\pm 1$ & $\pm6.6\cdot10^{-7}$ \\
    \hline
    \end{tabular}
    \caption{Uncertainties for flow quantities referred to in the paper}
    \label{tab:unc}
    
    \begin{tabular}{|m{0.7cm}|m{0.7cm}|m{0.7cm}|m{0.7cm}|m{0.7cm}|m{0.7cm}|m{0.7cm}|m{0.7cm}|m{0.7cm}|m{0.7cm}|m{0.7cm}|m{0.7cm}|}
    \hline
    \centering 1$\delta_2$ &  \centering 1.6$\delta_2$ &  \centering 2.2$\delta_2$ &  \centering 3$\delta_2$ &  \centering 5$\delta_2$ &  \centering 7$\delta_2$ &  \centering 9$\delta_2$ &  \centering 11$\delta_2$ &  \centering 15$\delta_2$ &  \centering 19$\delta_2$ &  \centering 27$\delta_2$ &  \multicolumn{1}{c|}{39$\delta_2$}  \\
    \hline
    \begin{center}
        \begin{tikzpicture}[baseline={(0,0.4)}]
            \fill[col1, draw=black] (0,0) -- (0.15,0.3) -- (-0.15,0.3) -- cycle;
        \end{tikzpicture}
    \end{center}
&
    \begin{center}
        \begin{tikzpicture}[baseline={(0,0.4)}]
            \fill[col2, draw=black] (0,0) -- (0.15,0.3) -- (-0.15,0.3) -- cycle;
        \end{tikzpicture}
    \end{center}
&
    \begin{center}
        \begin{tikzpicture}[baseline={(0,0.4)}]
            \fill[col3, draw=black] (0,0) -- (0.15,0.3) -- (-0.15,0.3) -- cycle;
        \end{tikzpicture}
    \end{center}
& 
    \begin{center}
        \begin{tikzpicture}[baseline={(0,0.4)}]
            \fill[col4, draw=black] (0,0) -- (0.15,0.3) -- (-0.15,0.3) -- cycle;
        \end{tikzpicture}
    \end{center}
&
    \begin{center}
        \begin{tikzpicture}[baseline={(0,0.4)}]
            \fill[col5, draw=black] (0,0) -- (0.15,0.3) -- (-0.15,0.3) -- cycle;
        \end{tikzpicture}
    \end{center}
&
    \begin{center}
        \begin{tikzpicture}[baseline={(0,0.4)}]
            \fill[col6, draw=black] (0,0) -- (0.15,0.3) -- (-0.15,0.3) -- cycle;
        \end{tikzpicture}
    \end{center}
&
    \begin{center}
        \begin{tikzpicture}[baseline={(0,0.4)}]
            \fill[col7, draw=black] (0,0) -- (0.15,0.3) -- (-0.15,0.3) -- cycle;
        \end{tikzpicture}
    \end{center}
&
    \begin{center}
        \begin{tikzpicture}[baseline={(0,0.4)}]
            \fill[col8, draw=black] (0,0) -- (0.15,0.3) -- (-0.15,0.3) -- cycle;
        \end{tikzpicture}
    \end{center}
& 
    \begin{center}
        \begin{tikzpicture}[baseline={(0,0.4)}]
            \fill[col9, draw=black] (0,0) -- (0.15,0.3) -- (-0.15,0.3) -- cycle;
        \end{tikzpicture}
    \end{center}
&
    \begin{center}
        \begin{tikzpicture}[baseline={(0,0.4)}]
            \fill[col10, draw=black] (0,0) -- (0.15,0.3) -- (-0.15,0.3) -- cycle;
        \end{tikzpicture}
    \end{center}
&
    \begin{center}
        \begin{tikzpicture}[baseline={(0,0.4)}]
            \fill[col11, draw=black] (0,0) -- (0.15,0.3) -- (-0.15,0.3) -- cycle;
        \end{tikzpicture}
    \end{center}
&
    \begin{center}
        \begin{tikzpicture}[baseline={(0,0.4)}]
            \fill[col12, draw=black] (0,0) -- (0.15,0.3) -- (-0.15,0.3) -- cycle;
        \end{tikzpicture}
    \end{center}
\\
    \hline
    \end{tabular}
    \caption{Colour legend for different roughness fetches applied to all plots in \S \ref{results}}
    \label{tb:legend}
\end{table}

The recovery of WSS with fetch length is shown more clearly in figure \ref{fig3}B. This plot shows the recovery of the friction coefficient measured at around 20 [ms$^{-1}$] with fetch length. This is the lowest flow speed at which the TBL seems to reach fully rough conditions and is thus used for the PIV measurements as well. The friction coefficient is plotted against the normalised fetch length, where $\delta_2$ is the boundary layer thickness at the balance location of the fully rough case with a fetch length of $\approx 39\delta_2$. This length scale was chosen instead of the most commonly used $\delta_1$ for two reasons. Firstly, having the recovery length as a function of the downstream TBL thickness removes all dependency on the type of surface upstream of the step change, making it applicable to more cases; secondly, the TBL thickness was measured using PIV above the balance to ensure consistency between the balance readings and the flow field above while no PIV measurement was taken upstream of the step change in any of the cases. For comparison, table \ref{tab:delta} lists the TBL thickness measured above the balance for all the different fetches. 
\begin{table}
    \centering
    \begin{tabular}{|r|c|c|c|c|c|c|c|}
    \hline
    $L \;\;[m]$ & $1\delta_2$ &  3$\delta_2$ &  5$\delta_2$ &  7$\delta_2$ &  9$\delta_2$ &  19$\delta_2$ &  39$\delta_2$  \\
    \hline
    $\delta_{99} \;[m]$ & $0.1202$ &  $0.1205$ &  $0.1219$ &  $0.1262$ &  $0.1265$ &  $0.1292$ &  $0.1495$  \\
    \hline
    \end{tabular}
    \caption{TBL thickness at the drag balance location of the cases tested with PIV, fetch length defined as a function of $\delta_2$ (the TBL thickness of the case with longest fetch).}
    \label{tab:delta}
\end{table}

Figure 2.4 in \citet{Li2020} shows a comparison of recovery lengths collated from previous studies, suggesting a wide range of recovery lengths from 1  to 10 $\delta_1$ for both smooth-to-rough and rough-to-smooth transitions. Our present results, shown in figure \ref{fig3}B, suggest a longer recovery length of at least $20\delta_2$. Secondly, although we expect the overshoot in $C_f$ immediately after transition (i.e. $C_f$ measured in shorter patch lengths, $1\delta_2-5\delta_2$), we observe that for $L> 5\delta_2$, the error in $C_f$ is within $\approx 10\%$ of the converged value. This type of error is to be expected when a shorter development length or computational domain is used.
\begin{figure}
\begin{subfigure}[t]{.5\textwidth}
\begin{tikzpicture}[scale=1]
\node at (0,0) {\includegraphics{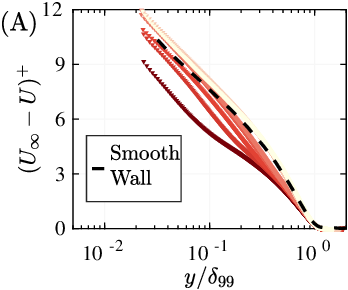}};
\end{tikzpicture}
\end{subfigure}\begin{subfigure}[t]{.5\textwidth}
\begin{tikzpicture}[scale=1]
\node at (0,0) {\includegraphics{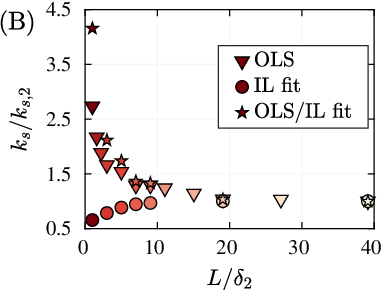}};
\end{tikzpicture}
\end{subfigure}
\caption{In (A) velocity defect plotted against $y$ normalised by $\delta_{99}$ as listed in table \ref{tab:delta} for each case. In (B) $k_s$ evolution, normalised by $k_{s,2}$, with fetch length $L$ scaled with $\delta_2$. $k_{s,OLS}$ calculated at  \href{https://cocalc.com/share/public_paths/68a2c17f4a61dc0cebfc9b50b813aec1a3021394}{\color{purple}{this link}}}
\label{fig4}
\end{figure}
Figure \ref{fig3}A can also be used to obtain the equivalent sand-grain height following the method proposed by \citet{Monty2016}. As briefly mentioned in \S1, this method assumes that the flow has already reached an equilibrium state and therefore employs \textit{outer layer similarity} from \citet{Townsend1965}, to obtain a relationship between $C_f$ at constant unit Reynolds number and $k_s$, via what the authors refer to as ``lines of constant length, $k_s/x$". These are shown in figure \ref{fig3}A as pink horizontal solid lines and, the intersection of these and the $C_f$ values at given $Re_x$, give us a way of calculating $k_s$ for different fetch lengths.

Before discussing the result of this operation, the outer layer similarity hypothesis from \citet{Townsend1965} was reproduced and is shown in figure \ref{fig4}A. From this plot, it can be seen that for shorter fetch lengths, velocity defect profiles do not collapse and hence do not conform to Outer Layer Similarity (OLS). This is to be expected as OLS is a measure of equilibrium with the boundary layer and for fetches lower than $\approx 10\delta_2$ equilibrium cannot be achieved due to the development of the internal layer. On the other hand, for the longer fetches, OLS can be observed as the profiles perfectly collapse onto smooth wall TBLs from $\approx0.4\delta$. The latter is the main conclusion from \citet{Townsend1965}, which means that the near wall region and anything that is associated with it should not affect the outer portion of the TBL. From our results, we can conclude that this indeed holds for the longest fetches tested. In the following analysis, we will see more in detail how the non-equilibrium conditions affect the prediction of $k_s$ values based on OLS and how this compares to the standard practice of calculating it from the roughness function $\Delta U^+$ where fully rough as well as equilibrium conditions are assumed.

In figure \ref{fig4}B we show the $k_s$ development with fetch length obtained using two methods. Firstly, 
the method from \citet{Monty2016} defined by the symbol \begin{tikzpicture}[baseline={(0,0.4)}]
\fill[col1, draw=black] (0,0.4) -- (0.15,0.7) -- (-0.15,0.7) -- cycle; \end{tikzpicture}; secondly, we used equation \ref{eq:velocitylaw} to fit logarithmic profiles to the velocity profiles near the wall, represented by the symbol \begin{tikzpicture}[baseline={(0,0.4)}] \filldraw[col1,draw=black] (0,0.55) circle (4pt);           
\end{tikzpicture}. Lastly, the symbol \begin{tikzpicture}[baseline={(0,0.4)}] \filldraw[col1,draw=black] (0,0.4) -- (0.15,0.5) -- (0.3,0.4) -- (0.24,0.55) -- (0.35,0.65) -- (0.2,0.65) -- (0.15,0.8) -- (0.1,0.65) -- (-0.05,0.65) -- (0.06,0.55) --cycle;  \end{tikzpicture} represents the ratio of the $k_s$ calculated with the two methods to offer a direct comparison of the respective values with increasing fetch length. Starting with the method from \citet{Monty2016} we observe that $k_s$ follows the same trend as $C_f$ i.e. overshooting its ``real" value for a certain surface at fixed Reynolds number and logarithmically approaching its true value with increasing fetch length. This plot shows how crucial it is to ensure sufficient fetch length for the WSS to recover to be able to treat $k_s$ as universal and use it as a length scale/modelling constant for rough wall TBLs. It can also be noted that the minimum fetch length for full WSS recovery is around $L\geq20\delta_2$, where $C_f$ becomes both Re and fetch length independent. We note here that this streamwise fetch might depend on the type of roughness and the extent of change in $k_s$ (from upstream to downstream). If the change is small, then, we expect the surface to reach equilibrium faster. Regardless, the results suggest that TBLs flowing over a change in wall texture with fetch lengths shorter than at least $5\delta_2$ ($\text{error} \geq 10\%$) will inevitably result in a significant overestimation of the roughness function and corresponding mean flow.

In figure \ref{fig4}B, we also see the trend of $k_s$ when calculated by fitting a logarithmic profile to the velocity profile in the near-wall region (i.e. below the inflection point), which is the point where the IL blends into the outer layer. This is done by applying equation \ref{eq:velocitylaw} to obtain the $k_s$ value that gives the best matching $U^+$ profile via an iterative procedure. As shown in this figure, the trend captured by this method is opposite to the one given by the previous one. Nonetheless, the fetch length at which we can infer equilibrium conditions after a step change does not change and is in full agreement between the two methods. Moreover, the converged value of $k_s$ for the longer fetch cases appears to be in perfect agreement as well. The challenge in using this method lies in accessing velocity measurements in the region close to the wall with enough resolution to fit a logarithmic profile and achieving a Reynolds number large enough to be able to distinguish the inflection point.

\begin{figure}
\begin{subfigure}[t]{.5\textwidth}
\centerline{\includegraphics{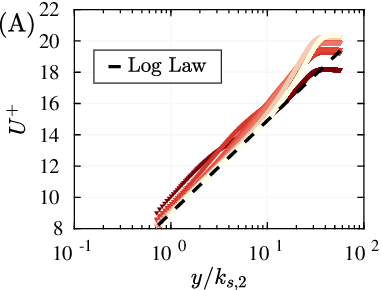}}
\end{subfigure}\begin{subfigure}[t]{.5\textwidth}
\centerline{\includegraphics{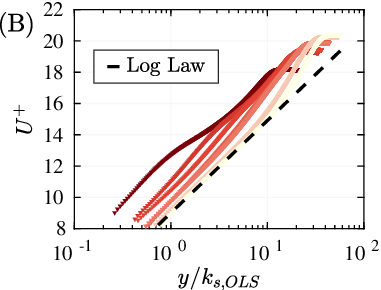}}
\end{subfigure}
\begin{subfigure}[t]{\textwidth}
\centering
\includegraphics{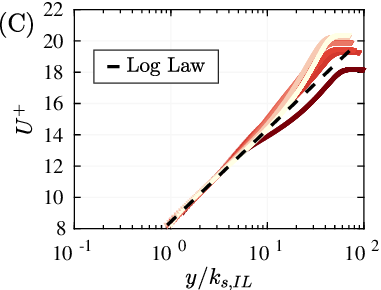}
\end{subfigure}
\caption{Velocity in viscous units plotted against the wall-normal coordinate normalised by (A) $k_{s,2}$ given by the longest fetch case, (B) $k_{s,OLS}$ shown with the symbol \begin{tikzpicture}[baseline={(0,0.4)}]
\fill[col1, draw=black] (0,0.4) -- (0.15,0.7) -- (-0.15,0.7) -- cycle; \end{tikzpicture} in figure \ref{fig4}B, and (C) $k_{s,IL}$ shown with the symbol \begin{tikzpicture}[baseline={(0,0.4)}]
\fill[col1, draw=black] (0,0.55) circle (4pt); \end{tikzpicture} in figure \ref{fig4}B}
\label{fig5} 
\end{figure}

Figure \ref{fig5} shows the streamwise-averaged mean flow profiles measured by PIV, taken above the FE drag balance and scaled by the friction velocity given by the balance measurement, where the black dashed line represents the log profile. In figure \ref{fig5}A, the wall-normal coordinate used to plot all the profiles is normalised by the fully-rough, equilibrium value of $k_s = k_{s,2}$, which is computed for the longest fetch case. Here, we can see that although the two longest fetches collapse onto the dashed line perfectly in the log region, the rest of the cases slowly diverge from it with the shortest fetch displaying a change in slope across the logarithmic domain of the TBL. This is clearly explained by the blending of the logarithmic regions from the upstream and downstream surface near the step-change in roughness. In the next plot, figure \ref{fig5}B, we used a $k_s$ value for each fetch case that attempts to include the effect of the internal layer development by computing it using the local $C_f$ value for each case. However, when using this method the profiles seem to diverge to a greater extent than using than $k_{s,2}$ value for all the cases. This can be explained by the equilibrium assumption made when employing the method described in \citet{Monty2016}. Lastly, in figure \ref{fig5}C, we used the $k_s$ value for each individual case given by fitting a logarithmic profile to the IL only as given in figure \ref{fig1}. Using this method we achieved a perfect match for all fetches below the inflection point, while above this point the shorter fetch profiles do not collapse onto the longer ones. This is because the $k_s$ value that models the IL region would inevitably fail in the outer region in cases of non-equilibrium such as a TBL after a step change in roughness. Therefore, in order to achieve a universal scaling we would have to make $k_s$ a function of $x$, by employing a different value for different fetches, and $y$, by using a different value below and above the inflection points where the IL is still developing. Finally, this method is only possible when a direct way of measuring drag is available as the drag given by the slope of the IL is not correct for short fetches. 

\section{Conclusions}
\label{conclusion}
The current paper aims to describe the outcome of an experimental campaign involving direct measurements of the WSS recovery after a step change in wall roughness with systematically increased fetch length. The results show that full WSS recovery is achieved $20\delta_2$ downstream of the step change, while previous studies employing indirect ways of measuring the WSS recovery predicted a full recovery between $1\delta_1-10\delta_1$. This difference is most likely due to the logarithmic nature of the WSS recovery. Therefore, even the smallest difference in WSS results in a significant difference in fetch length. We also show that the greatest change in WSS appears for fetch lengths between $1\delta_2-5\delta_2$, resulting in an error of $\leq 10\%$ of the converged WSS value when fetches $\geq 5\delta_2$ are used. 

Moreover, we have shown that the equivalent sand-grain height, $k_s$, given by the method adopted in \citet{Monty2016} cannot be used to scale or model the mean velocity profile of a TBL for fetches measuring less than $5\delta_2$, as this would inevitably result in a significant overprediction of the roughness effects and erroneous velocity profiles. This is due to the assumptions employed when deriving $k_s$, including fully-rough regime and equilibrium conditions in the TBL, which do not apply in the case of a TBL flowing past a step change with fetch length measuring less than $5\delta_2$. On the other hand, when fitting a logarithmic profile to the IL region, we can achieve a unique $k_s$ value for finite fetches that is able to scale/model the velocity profile below the inflection point and, by making $k_s$ vary in the wall-normal direction, we could be able to scale/model TBLs past step changes in roughness and their development to a greater extent.

A new way of modelling $k_s$, which takes into account both log regions of the internal boundary layer downstream of the transition and the outer layer (containing the flow history prior to the transition), could help with scaling/modelling streamwise varying rough wall TBLs. A correction factor between the $k_s$ trend with increasing fetch given by \citet{Monty2016} and the one given by fitting should also be developed in cases where high-resolution PIV at the right Reynold number near the wall is not viable. \\

\noindent \textbf{Acknowledgments}. The authors acknowledge funding from the Leverhulme Early Career Fellowship (Grant ref: ECF-2022-295), the European Office for Airforce Research and Development  (Grant ref: FA8655-23-1-7005) and EPSRC (Grant ref no: EP/W026090/1).\\

\noindent \textbf{Data Statement}. All data presented in this manuscript will be made publicly available upon publication. 

\bibliographystyle{jfm}
\bibliography{jfm-instructions}

\end{document}